\documentclass{article}

\def\ex{\exists}
\def\fa{\forall}
\def\ra{\rightarrow}

\begin{document}
\title{Skolemization in Simple
  Type Theory: the Logical and the Theoretical Points of View}

\author{Gilles Dowek\thanks{
Ecole polytechnique and INRIA, LIX, \'Ecole polytechnique, 91128 Palaiseau Cedex, France.
{\tt gilles.dowek@polytechnique.fr}}}

\date{}
\maketitle
\thispagestyle{empty}

Peter Andrews has proposed, in 1971, the problem of finding an analog
of the Skolem theorem for Simple Type Theory. A first idea lead to a
naive rule that worked only for Simple Type Theory with the axiom of
choice and the general case has only been solved, more than ten years
later, by Dale Miller \cite{Miller,miller87sl}. More recently, we have
proposed with Th\'er\`ese Hardin and Claude Kirchner \cite{HOL-ls} a
new way to prove analogs of the Miller theorem for different, but
equivalent, formulations of Simple Type Theory.

In this paper, that does not contain new technical results, I try to
show that the history of the skolemization problem and of its various
solutions is an illustration of a tension between two points of view
on Simple Type Theory: the {\em logical} and the {\em theoretical}
points of view.

\section{Skolemization}

\subsection{The Skolem theorem}

Let ${\cal T}$ be a theory in first-order predicate logic containing
an axiom of the form 
$$\fa x_1 ...\fa x_n \ex y~A$$
and ${\cal T}'$ be the theory obtained by replacing this axiom by 
$$\fa x_1 ...\fa x_n~((f(x_1, ..., x_n)/y)A)$$ 
where $f$ is a function symbol not used in ${\cal T}$.
Then, the Skolem theorem asserts that
the theory ${\cal T}'$ is a conservative extension of ${\cal
T}$ and, in particular, that one theory is contradictory if and
only if  the other is. 

\subsection{Extending Skolem theorem to Simple Type Theory}

The Skolem theorem plays a key role in automated theorem proving
because it permits to eliminate quantifier alternation in the
proposition to be proved, or refuted, and this alternation is often
delicate to manage.  This explains why, in 1971, seeking for a
generalization of the Resolution method to Simple Type Theory
\cite{Andrews71}, Peter Andrews has proposed the problem of finding an
analog of the Skolem theorem for Simple Type Theory.

Following the Skolem theorem for first-order predicate logic, we can
try to replace an axiom of the form
$$\fa x_1 ...\fa x_n \ex y~A$$
where $x_1$, ..., $x_n$ and $y$ are variables of type $T_1$, ...,
$T_n$ and $U$, by 
$$\fa x_1 ...\fa x_n~(((f~x_1~...~x_n)/y)A)$$ 
where $f$ is now a new constant of type $T_1 \ra ... \ra T_n \ra U$.
Unfortunately, the theory ${\cal T}'$ obtained this way is not always
a conservative extension of the theory ${\cal T}$. 
For instance, it is not possible to prove the proposition $\ex g~\fa
x~(P~x~(g~x))$ from the axiom $\fa x~\ex y~(P~x~y)$, 
because,
as proved again by Peter Andrews in 1972 \cite{Andrews72}, 
the axiom of choice 
$$(\fa x~\ex y~(P~x~y)) \Rightarrow (\ex g~\fa x~(P~x~(g~x)))$$ is not
provable in Simple Type Theory. But this proposition is obviously
provable from the skolemized form of this axiom: $\fa x~(P~x~(f~x))$.
Thus, although this naive skolemization can be used in Simple Type
Theory extended with the axiom of choice, it cannot be used in the
usual formulation of Simple Type Theory, without the axiom of choice.

A more restricted form of skolemization has been proposed in 1983 by
Dale Miller. In the Miller theorem, the Skolem symbol $f$ is not only
given the type $T_1 \ra ... \ra T_n \ra U$ but also the arity $\langle
T_1, ..., T_n, U \rangle$ and, unlike the usual symbols of a
functional type, the Skolem
symbols are not terms {\em per se}. To form a term with a Skolem
symbol $f$ of arity $\langle T_1, ..., T_n, U \rangle$, it is
necessary to apply it to terms $t_1$, ..., $t_n$ of type $T_1$, ...,
$T_n$, called the {\em necessary arguments} of the symbol $f$. This
reflects the intuition that, in a model of the negation of the axiom of
choice, we may have for each $n$-uple $t_1$, ..., $t_n$ the object
$(f~t_1~...~t_n)$, without having the function $f$ itself as an object. 

As it is not a term, the symbol $f$ cannot be used as a
witness to prove the proposition $\ex g~\fa x~(P~x~(g~x))$.  However,
this restriction is not sufficient, because, although the symbol $f$
cannot be used as a witness, the term $\lambda z~(f~z)$, can, yielding,
after normalization, the same result: $\fa x~(P~x~(f~x))$.  This
explains why Miller has introduced a second restriction: that the
variables free in the necessary arguments of a Skolem symbol
cannot be bound by a $\lambda$-abstraction, higher in a term
substituted for a variable in a quantifier rule.

With these two restrictions, Miller has been able to give a syntactic
proof of an analogous of the Skolem theorem: any proof using the axiom 
$$\fa x_1 ...\fa x_n~(((f~x_1~...~x_n)/y)A)$$ 
and whose conclusion does not use the symbol $f$ can be transformed
into a proof using the axiom  
$$\fa x_1 ...\fa x_n \ex y~A$$

\section{The logical and the theoretical point of view}

Simple Type Theory is otherwise known as {\em higher-order logic}. This
duality of names reveals that this formalism is sometimes seen as a
theory and sometimes as a logic.  As a theory, it should be compared
to other theories, such as arithmetic or set theory.  As a logic, to
other logics, and in particular to first-order predicate logic, that,
historically, is a restriction of it.

From the logical point of view, it is natural to try to {\em extend} the
theorems and algorithms known for first-order predicate logic, such as
the G\"odel completeness theorem, the Skolem theorem, the Resolution
method, ... to higher-order predicate logic. But, from the theoretical
point of view, it is more natural to try to express Simple Type
Theory as an 
axiomatic theory in first-order predicate logic and {\em apply} these
theorems and algorithms to this particular theory.

This tension between the logical and the theoretical point of view is
illustrated in a 1968 discussion between J. Alan Robinson and Martin
Davis \cite{RobinsonDavis,DavisRobinson}.  Robinson calls first-order
predicate logic: ``the restricted predicate calculus'' and
higher-order logic: ``the full predicate calculus'', while Davis,
replying that it is a simple matter to express Simple Type Theory as a
theory in first-order predicate logic, calls such a theory expressed in
first-order predicate logic a ``theory with standard formulation''.

The logical point of view has dominated the history of Simple Type
Theory: Henkin models, Higher-order resolution, ... have been designed
specifically for Simple Type Theory and not for a class of theories
formulated in first-order predicate logic, among which Simple Type
Theory is one instance. It is only recently that presentations of
Simple Type Theory in first-order predicate logic have shed a new
light on Henkin models, as anticipated by Martin Davis
\cite{DavisRobinson}, on higher-order unification, on proof search
algorithms, on cut elimination theorems, on functional interpretations
of constructive proofs, ... and on the Miller theorem.

Indeed, as shown in \cite{HOL-ls}, expressing Simple Type Theory as
a theory in first-order predicate logic allows to apply the Skolem
theorem and this way to reconstruct the Miller theorem.

\section{Simple Type Theory as a theory in first-order predicate logic}

We shall express Simple Type Theory as a theory in many-sorted
first-order predicate logic with equality. If a single-sorted formulation were
needed, the usual relativization method could be applied.

\subsection{Sorts}

As, in Simple Type Theory, functions and predicates are objects, the
sorts of the theory are not only the base types $\iota$ and $o$ of
Simple Type Theory but all its types: $\iota$, $o$, 
$\iota \ra \iota$, $\iota \ra o$, $o \ra o$, ... 

As
usual in many-sorted first-order predicate logic, terms are assigned a sort and
function and predicate symbols are assigned a tuple of sort called an 
{\em arity} or a {\em rank}.
If $f$ is a function symbol of arity 
$\langle T_1, ..., T_n, U \rangle$ and $t_1$, ..., $t_n$ are
terms of sort $T_1$, ..., $T_n$, then $f(t_1, ..., t_n)$ is a term of
sort $U$ and if $P$ is a predicate symbol of arity 
$\langle T_1, ..., T_n \rangle$ and $t_1$, ..., $t_n$ are
terms of sort $T_1$, ..., $T_n$, then $P(t_1, ..., t_n)$ is an atomic
proposition. 

\subsection{Symbols}

It is well-known that making a predicate $P$ an object requires to
introduce a copula $\in$ and write $a \in P$ what was previously
written $P(a)$. In the same way making a function $f$ an object
requires to introduce an application symbol $\alpha$ and write
$\alpha(f,a)$ what was previously written $f(a)$. 

In Simple Type Theory, we need a function symbol $\alpha_{T,U}$ of
arity $\langle T \ra U, T, U \rangle$ for each pair of sorts $T, U$
and a single predicate symbol $\varepsilon$ of arity $\langle o
\rangle$ to promote a term $t$ of sort $o$ to a proposition
$\varepsilon(t)$.

For instance, if $P$ is a term of sort $T \ra o$ and $t$ a term
of sort $T$, then the proposition usually written $P(t)$ is not
written $t \in P$, like in set theory, but
$\varepsilon(\alpha_{T,o}(P,t))$ where $P$ is first applied to $t$
using the function symbol $\alpha_{T,o}$ to build a term of sort $o$,
that is then promoted to a proposition using the predicate symbol
$\varepsilon$. In the same way, the proposition usually written $\fa P~(P
\Rightarrow P)$ is written $\fa p~(\varepsilon(p) \Rightarrow
\varepsilon(p))$.

Then, we need symbols to construct terms expressing functions and
predicates. Introducing the binding symbol $\lambda$ is not possible
in first-order predicate logic and we have to use a first-order
encoding of $\lambda$-calculus. A simple solution is to use the
combinators $S$ and $K$. Thus, for each triple of sorts $T, U, V$, we
introduce a constant $S_{T,U,V}$ of sort $(T \ra U \ra V) \ra (T \ra
U) \ra T \ra V$ and for each pair of sorts $T, U$, we introduce a
constant $K_{T,U}$ of sort $T \ra U \ra T$. Finally, we need similar
combinators to build terms of sort $o$. Thus, we introduce constants 
$\dot{=}_{T}$ of sort $T \ra T \ra o$, 
$\dot{\top}$ and $\dot{\bot}$ of sorts $o$, $\dot{\neg}$ of sort $o
\ra o$, $\dot{\wedge}$, $\dot{\vee}$ and $\dot{\Rightarrow}$ of sort
$o \ra o \ra o$, $\dot{\fa}_T$ and $\dot{\ex}_T$ of sort $(T \ra o)
\ra o$. Of course, some of these symbols are redundant and could be
defined from others using de Morgan's law. It is also possible to
define all these symbols from equality $\dot{=}_{T}$, following the
idea of Leon Henkin and Peter Andrews \cite{Henkin63,Andrews63}.

Indices may be omitted when they can be reconstructed from the context.

\subsection{Axioms}

Finally, we need axioms expressing the meaning of these
symbols. Besides the axioms of equality, we take the axioms
$$\fa x \fa y \fa z~(\alpha(\alpha(\alpha(S,x),y),z) =
\alpha(\alpha(x, z),\alpha(y,z)))$$
$$\fa x \fa y~(\alpha(\alpha(K,x),y) = x)$$
$$\fa x \fa y~(\varepsilon(\alpha(\alpha(\dot{=},x),y))
\Leftrightarrow (x = y))$$ 
$$\varepsilon(\dot{\top}) \Leftrightarrow \top$$
$$\varepsilon(\dot{\bot}) \Leftrightarrow \bot$$
$$\fa x~(\varepsilon(\alpha(\dot{\neg},x)) \Leftrightarrow \neg
\varepsilon(x))$$
$$\fa x \fa y~(\varepsilon(\alpha(\alpha(\dot{\wedge},x),y)) \Leftrightarrow (
\varepsilon(x) \wedge \varepsilon(y)))$$
$$\fa x \fa y~(\varepsilon(\alpha(\alpha(\dot{\vee},x),y)) \Leftrightarrow (
\varepsilon(x) \vee \varepsilon(y)))$$
$$\fa x \fa y~(\varepsilon(\alpha(\alpha(\dot{\Rightarrow},x),y))
\Leftrightarrow (\varepsilon(x) \Rightarrow \varepsilon(y)))$$
$$\fa x~(\varepsilon(\alpha(\dot{\fa},x)) \Leftrightarrow \fa y~\varepsilon(\alpha(x,y)))$$
$$\fa x~(\varepsilon(\alpha(\dot{\ex},x)) \Leftrightarrow \ex
y~\varepsilon(\alpha(x,y)))$$
To these axioms, we may add, as usual, the extensionality axioms, the
axiom of infinity, the description axiom and the axiom of choice.

To recall the choice we have made to express terms with combinators,
we call this theory {\em HOL-SK}.

\subsection{Properties}

An easy induction on the structure of $t$ permits to prove that for
each term $t$ there exists a term $u$ such that the proposition 
$$\alpha(u,x) = t$$
is provable. The term $u$ is often written $\hat{\lambda} x~t$.
In the same way,
an easy induction on the structure of $P$ permits to prove 
that for each proposition $P$ there exists a term $u$
such
that the proposition 
$$\varepsilon(u) \Leftrightarrow P$$ 
is provable.

This way, we can translate all the terms of the usual
formulation of Simple Type Theory with $\lambda$-calculus to terms of
HOL-SK and 
prove that if 
$t$ is a term of sort $o$ and $t'$ its translation, then $t$ is
provable in the usual formulation of Simple Type theory if and only if
$\varepsilon(t')$ is provable in HOL-SK. However, this theorem
requires that the extensionality axioms are added to both theories, because
the combinators $S$ and $K$ do not simulate $\lambda$-calculus exactly,
but only up to extensionality.

\section{Skolemization}

As a consequence of the Skolem theorem, we get that 
if ${\cal T}$ is a theory containing an axiom of the form 
$$\fa x_1 ...\fa x_n \ex y~A$$
where $x_1$, ..., $x_n$ and $y$ are variables of sorts $T_1$, ...,
$T_n$ and $U$, 
and ${\cal T}'$ is the theory obtained by replacing this axiom by 
$$\fa x_1 ...\fa x_n~((f(x_1, ..., x_n)/y)A)$$ then $\mbox{HOL-SK}
\cup {\cal T}'$ is a conservative extension of $\mbox{HOL-SK} \cup
{\cal T}$.  

The symbol $f$ is a function symbol of arity $\langle T_1, ..., T_n, U
\rangle$ and not a constant of sort $T_1 \ra
... \ra T_n \ra U$, first because the Skolem theorem for first-order
predicate logic introduces a function symbol and not a constant and
then because it ignores the internal structure of sorts. Thus, the
symbol $f$ alone is not a term, and the term obtained by applying the
symbol $f$ to the terms $t_1$, ..., $t_n$ is the term $f(t_1, ...,
t_n)$ and not the ill-formed term $\alpha(...\alpha(f,t_1), ...,
t_n)$. In contrast, when the sort $U$ has the form $V \ra W$, the term
$f(t_1, ..., t_n)$ can be further applied to a term $t_{n+1}$ using
the application symbol: $\alpha(f(t_1, ..., t_n),t_{n+1})$.

In short, the Skolem symbols are not at the level of the symbols $S$
or $K$, but at the level of the symbols $\alpha_{T,U}$.

We prove this way an analogous of the Miller theorem for HOL-SK. For
this formulation Miller's second condition vanishes as there is no
binder $\lambda$. All that remains is the first condition: the fact
that Skolem symbols must be applied.  Notice however that the term
$\hat{\lambda} x~t$ cannot be defined for all terms $t$ containing the
symbol $f$, but only those that do not have an occurrence of the
variable $x$ in an argument of the symbol $f$.

An advantage of expressing Simple Type Theory as a theory in
first-order predicate logic is that the proof of the Miller theorem is
simplified as it is then proved as a consequence of the Skolem
theorem. Moreover this shows that arities are not a feature of the
Skolem symbols only, but that all function symbols have arities, in
particular the symbols $\alpha_{T,U}$ and the Skolem symbols.

\section{Miller's second condition}

We may wonder if it is possible to go one step further and reconstruct
Miller's second condition as a consequence of the Skolem theorem for
first-order predicate logic. As we shall see, this is possible but
this requires to use a more precise first-order encoding of the
$\lambda$-calculus: the $\lambda$-calculus with nameless dummies
introduced by Nicolaas de Bruijn \cite{deBruijn72}.

\subsection{De Bruijn indices}

In a $\lambda$-term, we may add, to each occurrence of a bound variable, 
a natural number expressing the number of abstractions
separating the occurrence from its binder. For instance, adding 
indices to the term
$$\lambda x \lambda y~(x~\lambda z~(y~x~z))$$
yields 
$$\lambda x \lambda y~(x^2~\lambda z~(y^2~x^3~z^1))$$
Indeed, the index of the occurrence of the variable $z$ is $1$ because
the binder $\lambda z$ is just one level above in the term, while the
index of the second occurrence of the variable $x$ is $3$ as the
the binder $\lambda x$ is three levels above. Once the indices are
added this way, the names of the bound variables are immaterial and can
be dropped. We thus get the term 
$$\lambda  \lambda~(\_^2~\lambda ~(\_^2~\_^3~\_^1))$$

Notice that the term 
$\lambda x \lambda y~(x~\lambda z~(y~x~z))$
is closed but that its subterm 
$\lambda y~(x~\lambda z~(y~x~z))$
contains a free variable $x$. 
In the same way, in the term
$\lambda \lambda~(\_^2~\lambda ~(\_^2~\_^3~\_^1))$
all the de Bruijn indices refer to a binder, but its subterm 
$\lambda~(\_^2~\lambda ~(\_^2~\_^3~\_^1))$
contain two ``free indices'', that exceeded the number of binders
above them, and correspond to the former occurrences of the variable $x$. 

The $\lambda$-calculus with de Bruijn indices is a first-order language
 as,
once variables names have been dropped, the
symbol $\lambda$ is not a binder anymore. This language is formed with
an infinite number of constants $\_^1, \_^2, \_^3, ...$, a binary function
symbol $\alpha$ for application and a unary function symbol $\lambda$.
Closed terms, such as $\lambda~(\_^2~\lambda ~(\_^2~\_^3~\_^1))$ or
$\_^1$, may contain free indices, exceeding the number of binders
above them.

\subsection{Sorts}

When type-checking an open $\lambda$-term {\em \`a la} Church, {\em i.e.}
with explicitly typed bound variables, for instance
$$\lambda y_{((\iota \ra \iota) \ra \iota) \ra \iota \ra
\iota}~(x~\lambda z_{\iota}~(y~x~z))$$ it is necessary to have a
context defining the type of the free variables such as $x$. Indeed,
if $x$ is assigned the type $(\iota \ra \iota) \ra \iota$, then the
term is well-typed, but not if it is assigned, for instance, the type
$\iota$.

In the same way, when typing a term that may contains free de Bruijn
indices, even if this term is closed ({\em i.e.} does not contain
named variables), we need a context defining the types of the de
Bruijn indices exceeding the number of binders above them. 
This context is a finite list of types: the type of 
the indices exceeding by 1 the number of binders above them, that of the
indices exceeding by 2 the number of binders above them, ... 
For instance the term
$$\lambda_{((\iota \ra \iota) \ra \iota) \ra \iota \ra
  \iota}~(\_^2~\lambda_{\iota}~(\_^2~\_^3~\_^1))$$ 
has type 
$(((\iota \ra \iota) \ra \iota) \ra \iota \ra \iota) \ra \iota$ in the
context  $[(\iota \ra \iota) \ra \iota]$. 

Thus, even if it is closed, a term $t$ can be assigned a type $T$,
only relatively to a context $\Gamma$. In other words, it can be
absolutely assigned an ordered pair formed with a context $\Gamma$ and
a type $T$. We write such a pair $\Gamma \vdash T$.

When we express Simple Type Theory as a first-order theory, using the
$\lambda$-calculus with de Bruijn indices as a first-order encoding of
the $\lambda$-calculus, the sort are not just the simple types, like
in HOL-SK, but such pairs $\Gamma \vdash T$ formed with a list
$\Gamma$ of simple types and a simple type $T$.

In this formulation, the quantified variables, in contrast to the
$\lambda$-abstracted ones, are not replaced by de Bruijn indices but,
as in all theories expressed in first-order predicate logic, they are
kept as standard named variables and they are assigned a sort of the
form $\vdash T$, where the context is the empty list.

The last step of the construction of this theory would be to introduce
explicit substitutions, hence its name {\em HOL-$\lambda \sigma$}. I
do not want to go into these details here and the interested reader
can refer to \cite{HOL-ls}. But, I want to insist on two
points. First, the fact that this theory is {\em intentionally}
equivalent to the usual presentation of Simple Type Theory with
$\lambda$-calculus, {\em i.e.} even if the extensionality axioms are
not assumed. Second, that, as explained above, the sorts are pairs
$\Gamma \vdash T$, {\em i.e.} that they contain scoping
information. In particular terms of a sort $\vdash T$, with an empty
context, are de Bruijn-closed, {\em i.e.} they do not contain indices
that may be bound higher in the term by a $\lambda$-abstraction.

\subsection{Skolemization}

As a consequence of the Skolem theorem, we get that 
if ${\cal T}$ is a theory containing an axiom of the form 
$$\fa x_1 ...\fa x_n \ex y~A$$
where $x_1$, ..., $x_n$ and $y$ are variables of sort $\vdash T_1$, ...,
$\vdash T_n$ and $\vdash U$, 
and ${\cal T}'$ is the theory obtained by replacing this axiom by 
$$\fa x_1 ...\fa x_n~((f(x_1, ..., x_n)/y)A)$$ then
$\mbox{HOL-$\lambda \sigma$} \cup {\cal T}'$ is a conservative
extension of $\mbox{HOL-$\lambda \sigma$} \cup {\cal T}$.  The symbol
$f$ is a function symbol of arity $\langle \vdash T_1, ..., \vdash
T_n, \vdash U \rangle$. Thus, not only $f$ is a function symbol, but a
function symbol whose arguments must not contain indices
that may be bound by a $\lambda$-abstraction higher in a term. We get
this way exactly Miller's conditions. The first is rephrased as the
fact that $f$ is a function symbol and the second as the fact that the
arguments of these function symbol must have a sort with an empty
context.

We obtain this way an alternative proof of the Miller theorem for
HOL-$\lambda \sigma$, as consequence of the Skolem theorem. 

A difficult point in the Miller theorem is the discrepancy between the
general formation rules for the terms and the propositions and the
more restricted ones for the terms substituted for variables in
quantifier rules. As we have seen,
the variables free in the necessary arguments of
the Skolem symbols cannot be bound in a term substituted for a
variable. But this restriction does not apply to the terms and
propositions in general, because, if it did, the skolemized axiom
itself would not be well-formed, as the arguments of the Skolem symbol
in this skolemized axiom are universally bound variables.

In HOL-$\lambda \sigma$ the situation is slightly different. The
formation rules for the terms substituted for variables in quantifier
rules are the same as those of the other terms. But the $\lambda$-bound
variables and the quantified variables are treated differently. As we
have seen, the $\lambda$-bound variables are replaced by de Bruijn
indices, but the quantified variables are kept as named variables, as
in all theories expressed in first-order predicate logic.  Thus the
$\lambda$-bound variables cannot appear in the arguments of a Skolem
symbol but the quantified variables can and, this way, the skolemized
axiom is well-formed.

\section{From the theoretical point of view}

With this example of skolemization, we have seen two advantages of the
theoretical point of view on Simple Type Theory. First the proofs of
the theorems are simplified, because we can take advantage of
having already proved similar theorems for first-order predicate
logic. Then some difficult points of these theorems are explained.

We can mention several other theorems and algorithms that have been
simplified and explained this way. The Henkin completeness theorem,
already mentioned in 1968 by Martin Davis, independence
results, in particular of the extensionality axioms, cut
elimination theorems, both model-based ones and reduction-based ones,
proof search algorithms and the functional interpretation of
constructive proofs.

This succession of points of view on Simple Type Theory is itself an
example of a common back and forth movement in the development of
science: first, new objects and new results are discovered, breaking
with the old framework, they are supposed not to fit in. Then, after a
possible evolution of the general framework, these new objects are
integrated back. We may try to understand what kind of
evolution of first-order predicate logic, the integration of Simple
Type Theory requires or suggests.

The first evolution is the shift from the single-sorted first-order
predicate logic to the many-sorted one. Although it is always possible
to relativize a many-sorted theory to a single-sorted one, the natural
framework to express Simple Type Theory is many-sorted first-order
predicate logic. It is interesting to see that early papers on
many-sorted first-order predicate logic, {\em e.g.} \cite{Wang},
already motivate the introduction of many-sorted first-order predicate
logic by the will to express Simple Type Theory in it.

A second evolution is to take into account the possibility to
consider terms and propositions up to reduction. The axioms of HOL-SK
(and those of HOL-$\lambda\sigma$) can easily be transformed into
rewrite rules:
$$\alpha(\alpha(\alpha(S,x),y),z) \longrightarrow \alpha(\alpha(x, z),\alpha(y,z))$$
$$\alpha(\alpha(K,x),y) \longrightarrow x$$
$$\varepsilon(\alpha(\alpha(\dot{=},x),y)) \longrightarrow x = y$$ 
$$\varepsilon(\dot{\top}) \longrightarrow \top$$
$$\varepsilon(\dot{\bot}) \longrightarrow \bot$$
$$\varepsilon(\alpha(\dot{\neg},x)) \longrightarrow \neg \varepsilon(x)$$
etc. and identifying equivalent propositions is more natural than
keeping these axioms as such, exactly like in the usual formulation of
Simple Type Theory, $\beta$-normalizing terms and propositions is more
natural than keeping $\beta$-conversion as an axiom. 
This idea has lead to the development of
{\em Deduction modulo}, that again was initially motivated by the will
to express Simple Type Theory in it.

A third evolution is the introduction of binders in first-order
predicate logic. Although this has been a hot topic recently, it
is fair to say that we have not yet a completely satisfactory
extension of first-order predicate logic with such binders.

To conclude, I want to mention the influence of Peter Andrews' work,
course and book \cite{Andrews86}, on the emergence of the theoretical
point of view on Simple Type Theory (although I do not attribute any
point of view to anyone except myself). First, it is striking that in
his course, Peter Andrews compared Simple Type Theory not only to
first-order predicate logic but also to set theory in particular that
he insisted on the fact that both theories are restriction of the
naive, inconsistent, set theory with full comprehension. More
technically, the alternative characterization of Henkin models, given
by Peter Andrews \cite{Andrews72} showed the way out of Henkin
mysterious conditions that all $\lambda$-terms must have a denotation
to a standard condition that the axioms $S$ and $K$ must be valid and
thus to the idea that Henkin models were just models of some theory
expressed in first-order predicate logic.

\section{Acknowledgements}

The author wants to thank Claude Kirchner and Dale Miller for helpful
comments on a previous version of this paper.

\bibliographystyle{plain-casefix} 
\bibliography{andrews}

\end{document}